\documentclass{autart}

\usepackage{graphics}
\usepackage{epsfig}
\usepackage{amssymb}

\def\cal{\fam2 }

\begin{document}

\begin{frontmatter}



\title{Cosmogenic Neutrinos from Ultra-High Energy Nuclei}

\author{M. Ave$^{1,4}$},\author{N. Busca$^{1,3}$},\author{A. V.\ Olinto$^{1,2,4}$}, \author{A.A. Watson$^5$}, \author{T. Yamamoto $^{1,4}$}

\address{$^1$Kavli Institute of Cosmological Physics}
\address{$^2$Department of Astronomy and Astrophysics}
\address{$^3$Department of Physics}
\address{$^4$ Enrico Fermi Institute, \\
The University of Chicago, 5640 S. Ellis, Chicago, IL60637, USA}
\address{$^5$Department of Physics and Astronomy\\
 University of Leeds, Leeds LS2 9JT ,UK}%

\begin{abstract}
We calculate the flux of neutrinos generated by the propagation of ultra-high energy  iron over cosmological distances and show that even if ultra-high energy cosmic rays are composed of heavy nuclei, a significant flux of high-energy neutrinos should be present throughout the universe. The resulting neutrino flux has a new peak at $\sim 10^{14} eV$ generated by neutron decay and reproduces the double peak structure due to photopion production at higher energies ($\sim 10^{18}$ eV). Depending on the  maximum energy and cosmological evolution of extremely high energy cosmic accelerators the generated neutrino flux can be detected by future experiments.
\end{abstract}

\begin{keyword}

\PACS 
\end{keyword}
\end{frontmatter}


\section{Introduction}

The  possibility of discovering new astrophysical phenomena through the observation of high energy neutrinos  has inspired a number of recent experimental efforts. Unlike photons and baryons, high energy neutrinos can traverse the universe without suffering energy losses and thus provide a probe into high energy phenomena throughout the entire universe. The study of high energy cosmic neutrinos also offers the opportunity to test particle interactions at energies well above terrestrial accelerators.  However, the fluxes of known astrophysical sources are expected to be challenging  low and uncertain. It is often claimed that the only {\it guaranteed} source of very high energy neutrinos are ultra-high energy cosmic rays (UHECRs). 

Ultra-high energy cosmic ray protons originating in extragalactic sources interact with the cosmic microwave background (CMB) radiation and generate neutrinos through the photopion production and subsequent pion decay  \cite{BerZat69}. These neutrinos are  called cosmogenic, photopion, or even GZK neutrinos as the photopion production responsible for their generation also induces a strong feature in the cosmic ray spectrum known as the GZK  cutoff  (after Greisen, Zatsepin, and Kuzmin \cite{Greisen,ZK}).  This cosmogenic neutrino flux has been extensively studied \cite{WTW,Stecker73,BerSmi75,BerZat77,Stecker79,HS85,HSW86,SDSS,PJ00,YT93,ESS01} since it represents the best hope for detectability at high energies. However, the cosmogenic flux is not {\it guaranteed} because the composition and source of cosmic rays at the extremely high energies is unknown (for reviews on UHECRs see, e.g.,  \cite{cronin04,nagwat00,BatSigl00,Olinto00,Gaisser00c}).  While there is a common assumption that protons dominate at the highest energies, hard experimental evidence is lacking at energies above $\sim 10^{19}$ eV. 
Recent composition studies indicate the dominance of  heavier primaries between $10^{17} $eV  and $10^{18}$ eV  followed by a lighter composition up to  $\sim 10^{19.4}$ eV \cite{hirescomp}. For a detectable cosmogenic neutrino flux, protons are assumed to be the main cosmic ray  primaries up to energies $\sim 10^{21} $ eV. 

Here we show that even if the highest energy cosmic rays are heavier nuclei, such as iron, a significant flux of ``propagation'' neutrinos should exist \cite{cris04,leeds04}. 
Heavy nuclei are attractive as UHECR primaries because they can be more easily accelerated than protons  in  astrophysical sources (their  rigidity is smaller at the same energy), and they are more easily isotropized by magnetic fields.  The absence of correlations between the arrival direction of UHECRs and candidate sources can be explained by the larger deflection of heavy nuclei in intergalactic magnetic fields.  The interaction  of  heavy nuclei with cosmic backgrounds as they  propagate through intergalactic space also generates a feature in the UHECR spectrum \cite{Greisen,ZK,stecker_photodisent,epele,stecker_update,toko,Sigl,anchod} that should be detectable by future experiments, such as the Pierre Auger Observatory and EUSO.
 
\section{Interaction of nuclei with background photons}

\begin{figure}
\epsfig{file=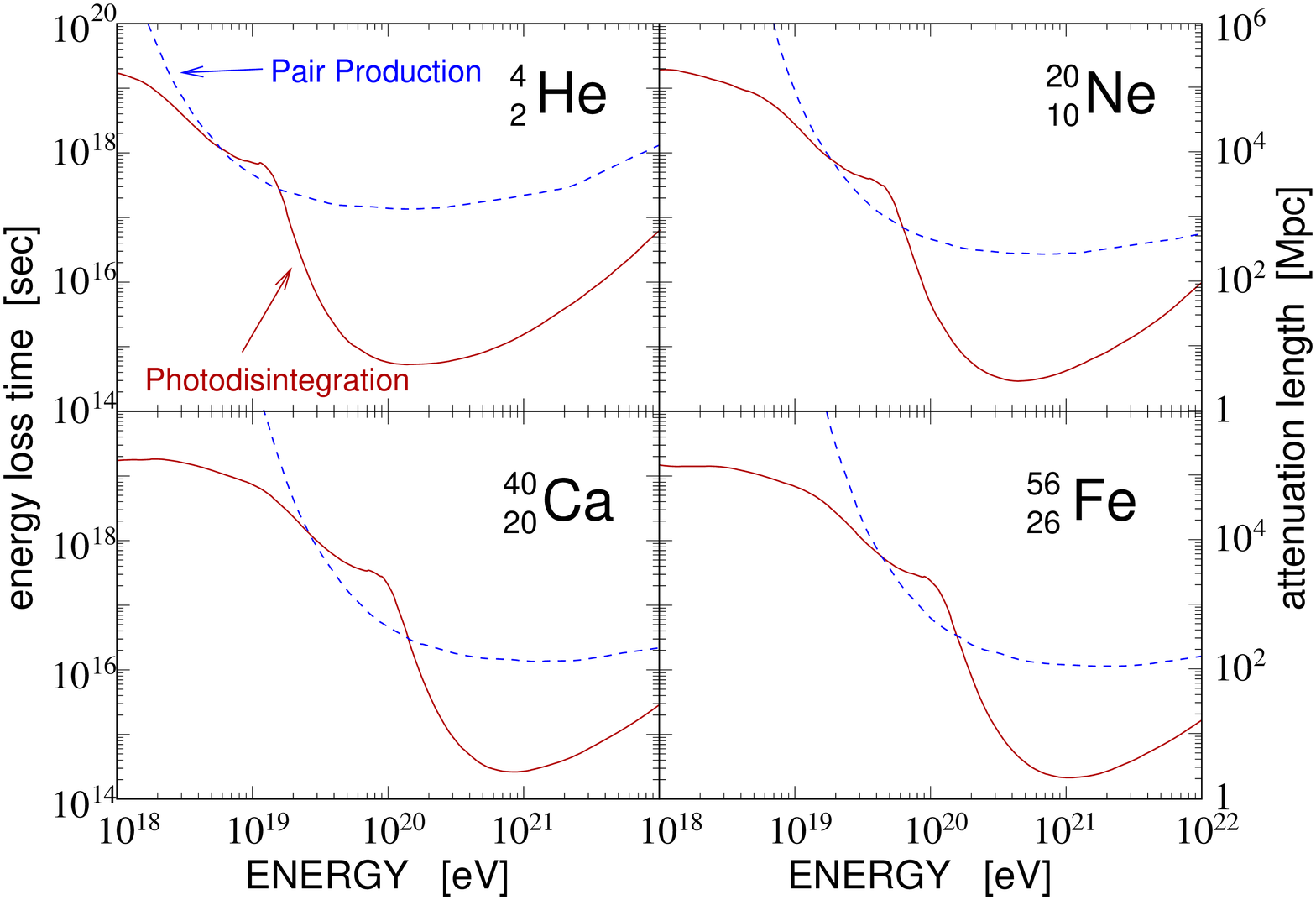,width=1.0\linewidth,height=0.8\linewidth }
\caption{Energy loss time of different mass nuclei as a function of energy. 
Solid line is that of single-nucleon emission by photodisintegration 
and dashed line is pair production \cite{steck1}.}
\label{fig:f1}
\end{figure}

The most relevant interactions between cosmic background photons  and ultra-high energy nuclei are pair production in the field of the nuclei, photodisintegration of the nucleus, and photopion production.  
Pair production has a threshold of 2$m_e~c^2$, in the rest frame of the nucleus, and is proportional to $Z^2/A$ for a nucleus with charge $Ze$ and mass number $A$. Figure \ref{fig:f1} shows characteristic time, $\tau$, for the energy loss due to pair production on the cosmological backgrounds (dashed-line) as a function of energy for $^4$He, $^{20}$Ne, $^{40}$Ca, and $^{56}$ Fe, where
\begin{equation}
\frac{1}{\tau}=\frac{1}{E}\frac{dE}{dt} \  .
\end{equation}
 
In  Figure \ref{fig:f1}, we also show the timescales for photodisintegration of each nuclei by plotting the  single-nucleon emission process times (solid lines). Photodisintegration of a nucleus into lighter nuclei due to the interaction with background photons is dominant over pair production losses over most of the relevant energy range.  
 In terms of background photon energy  in the rest frame of the nucleus, $\epsilon^\prime$, there are two important contributions.  The first comes from the  low energy range $\epsilon^\prime \lesssim 30$ MeV, in the Giant Dipole Resonace region, where the emission of one or two nucleons dominates. The second contribution comes from energies between 30 and 150 MeV, where multi-nucleon energy losses are involved. In this paper, we calculate this process by a Monte-Carlo method according to
 \cite{stecker_photodisent,steck1}.  These authors parametrized the total cross section $\sigma(\epsilon)$ as a function of photon energy in the rest frame of the nucleus.  Then the probability of photodisintegration per unit length $R$ is calculated by the following equation:
\begin{equation}
R=\int_0^{\infty} n(\epsilon) \left[ \int_0^{2\gamma\epsilon}
	\frac{\sigma(\epsilon^\prime) \frac{\epsilon^\prime}{\epsilon}
	d \epsilon^\prime}{2 \gamma \epsilon} \right]
	d \epsilon \ \ ;
\end{equation}
where $\epsilon$ and $\epsilon^{\prime}$ are the energies of the photon in the lab frame and in the nucleus rest frame respectively, $n(\epsilon)$ is the differential number density of photons including the 
CMB and the infrared background (IRB) \cite{steck1,infrared},  and $\gamma$ 
 is the Lorentz factor of the nucleus. The term inside the bracket corresponds to the angle-averaged cross-section for a photon of energy $\epsilon$. 

Photodisintegration by  the CMB is maximized when $\gamma \sim 10^{10}$. In general, one or a few nucleons and a single lighter nucleus are emitted by this interaction. The Lorentz factor is conserved at each photodisintegration interaction, though it is reduced by pair-production. In the case of iron primaries, photodisintegration is most efficient at  energies  $\sim 10^{21}$ eV, yielding a value of $\tau = 2\times 10^{14}$ s (2 Mpc). At lower energies, the contribution of the IRB is more important.  Because the IRB density is much lower than the CMB density, $\tau$ increases rapidly at lower energies.  
For an iron nucleus of $10^{20}$ eV, $\tau \sim 2.5\times10^{17}$s (2.5Gpc).  

The third process relevant for the propagation of nuclei is the photopion production off CMB photons.  Photopion production in nuclei occurs at energies above  $ E_{\gamma \pi} \simeq A \  6 \times 10^{19}$ eV, i.e., $\sim 3 \times 10^{21}$ eV for iron nuclei.  At these energies, iron nuclei photodisintegrate completely in $\sim 4$ Mpc. Only iron nuclei with energies above $6 \times10^{21}$ eV  have photopion interaction lengths comparable with that of complete photodisintegration. For the energy range we consider, the neutrino emission due to photopion interactions of the primary nucleus is insignificant when compared to the 
the photopion interactions of the secondary nucleons emitted by the nuclei lead to a significant neutrino flux.

Secondary nucleons emitted by the propagating  nuclei suffer energy losses through pair production and photopion proccesses. The interaction length, $\lambda(E)$,  for photopion production is given by
\begin{equation}
{1\over\lambda(E)} = {1\over8\pi E^2} \int_{\epsilon_{th}}^{\infty}d\epsilon
{n(\epsilon)\over\epsilon^2}\int_{s_{th}}^{s_{max}}ds(s-m_N^2)\sigma_{N\gamma}(s) \ \ ,
\label{lambda}
\end{equation}
where $n(\epsilon)$ is the photon number density per unit energy per
unit volume for a blackbody spectrum with temperature T=$2.7$ K, and $\sigma$ is the
cross section for photopion production given in \cite{SOPHIA}.

For a nucleon, $N$, interacting with the CMB, we consider only the $\Delta$-resonance  interaction channel for the photopion production: $N+\gamma\rightarrow\Delta\rightarrow N+\pi$.  As a result of the interaction, a pion and a nucleon are produced. From isospin considerations, if the incoming nucleon is a proton (neutron), the outgoing nucleon has probability 2/3 to be a proton (neutron) and 1/3 probability to be a neutron (proton). Approximately 20 \% of the energy is carried by the pion. Neutral pions decay into photons while charged pions decay into $\mu$ (or $\bar\mu$) and $\bar\nu_{\mu} (\nu_{\mu})$. The muon carries 80\% of the energy while the neutrino (or antineutrino) carries 20\% from the free two body decay. The muons further decay into $e^{\pm}$ and $\nu_e (\bar\nu_e)$ and $\sim$ 30\% of the energy is carried by each particle.

\section{Neutrino Yields from Single Sources}

We calculate the spectra of particles arising from the propagation of high-energy nuclei by combining a Monte-Carlo simulation of neutrino yields due to nuclei propagation from single sources folded into an uniformly distributed source distribution. The single source neutrino yield is simulated using an adaptation of the code developed by \cite{toko} up to a distance of 500 Mpc. We inject 3000 iron nuclei for each of 30 energy bins between 10$^{19}$ eV and 10$^{22}$ eV.
For this energy range, iron nuclei are completely photodisintegrated at distances $\lesssim 300$ Mpc. We ignore magnetic fields in this work, but it is clear that their effect is to shorten this distance considerably. 

The propagation of iron nuclei  from the source generates a number of daughter nuclei and nucleons. For each nucleus we consider every photodesintegration process involving the emission of one or more nucleons.  The interaction length is calculated for each photodisintegration process and a propagation distance is calculated for each channel by
\begin{equation}
X_{dist}=-\lambda_{nuclei}(E_0)~ln(\psi) \ , \\
\label{xdist}
\end{equation}
where $\lambda_{nuclei}(E_0)$ is the interaction length for a given
 channel at energy $E_0$ and $\psi$ a random number between 0 and 1. The channel with smallest $X_{dist}$ is taken, and the nuclei are propagated in steps of 200 kpc to their interaction point. In each step  the energy is updated to include continuous energy losses by pair production and adiabatic losses. Adiabatic losses do not play an important role due to the small photodisintegration distances ($<$ 500 Mpc).   When the particle reaches its interaction point, we allow the interaction to happen with a
 probability given by the ratio of the interactions lengths for the energy before and after propagation to $X_{dist}$ of the chosen channel following \cite{stanev00}.

Protons and neutrons are propagated in a similar way to nuclei. In the case of neutrons, photopion production competes with neutron decay at energies larger than $4 \times 10^{20}$ eV. When neutrons decay,  the energy of the produced proton, electron, and electron antineutrino is calculated using free three-body decay.  Photopion interactions produce a nucleon and a pion that  further decays producing neutrinos or photons.  The propagation terminates when particles reach the
most distant ``observation'' sphere, decay, or get older than the age of the universe.

\begin{figure}
\epsfig{file=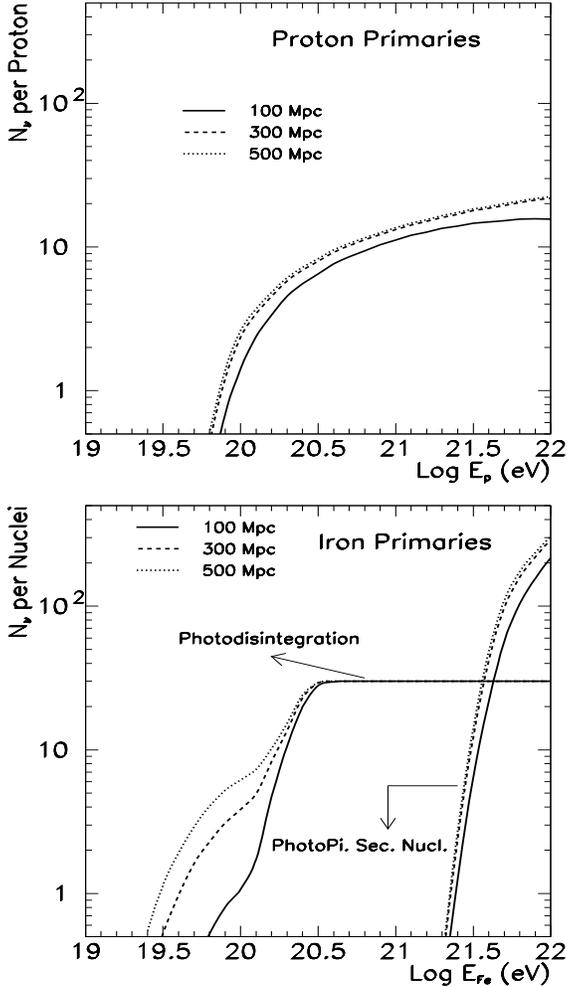,width=1.0\linewidth,height=1.7\linewidth }
\caption{Mean number of neutrinos produced by primary protons (left) and iron (right) after
 a progagation distance of 100 (solid), 300 (dashed) and 500 Mpc (dotted), as a function of the initial energy of the primary. The dashed lines in the case of iron correspond to neutrinos produced in the decay of neutrons emitted by the nuclei through photodisintegration.}
\label{numneu}
\end{figure}

Figure \ref{numneu} shows the mean number of neutrinos produced per primary,  for  propagation distances of 100, 300, and 500  Mpc, for proton primaries (left) and iron primaries (right) as a function of the primary's energy.
The neutrino yield is fully developed for a source distance of 300 Mpc.  For  iron primaries, solid lines correspond to neutrinos produced in the decay of pions from photopion interactions, while dashed lines correspond to neutrinos produced by the decay of neutrons from the disintegrated nuclei.  For energies above $\sim 4 \times10^{20}$ eV, iron nuclei completely  photodisintegrate over $\sim $ 10 Mpc (see fig. \ref{lambda}). In the disintegration of  iron,  30 neutrinos are produced in the decay of 30 neutrons. Above this energy, no additional neutrinos are produced until the threshold for photopion production is reached. For larger energies, the number of neutrinos produced increases rapidly due to the photopion production of 56 nucleons per iron nucleus.

\begin{figure}
\epsfig{file=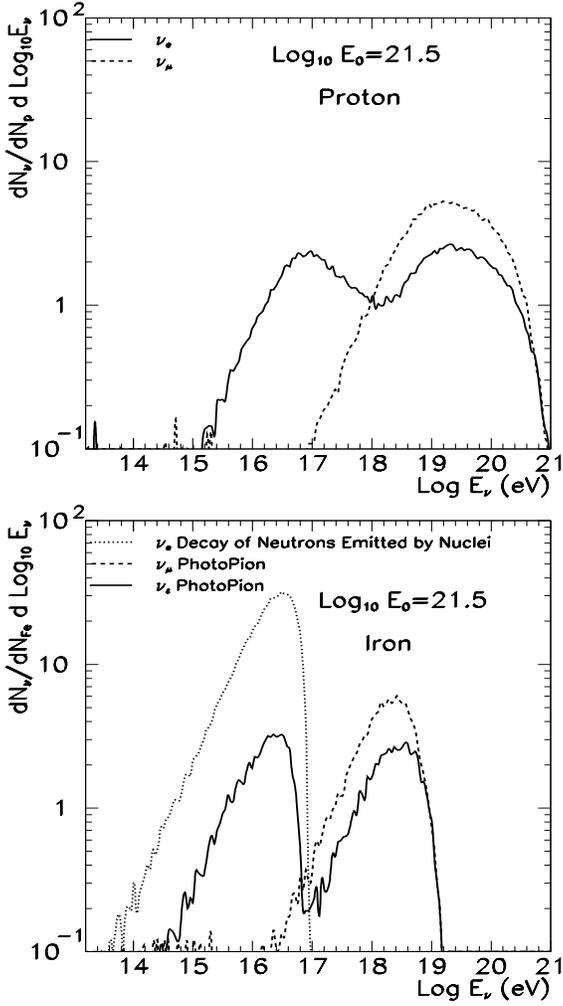,width=1.0\linewidth,height=1.7\linewidth }
\caption{Neutrino yield for proton (left) and iron (right). The energy considered is 
 10$^{21.5}$ eV, and the propagation distance 300 Mpc. Different lines indicate the
 different origin of the neutrinos: the solid lines are $\nu_e$+ $\bar\nu_e$ and the dashed lines are $\nu_\mu$+ $\bar\nu_e$ all  produced by photopion interactions. The dotted line is for $\bar\nu_e$ produced in the decay of neutrons emitted by the nuclei in photodisintegration processes.}
\label{dndloge}
\end{figure}

Figure \ref{dndloge} shows the fluxes of electron and muon neutrinos produced by protons (left) and iron (right) after propagation over a distance of 300 Mpc. The maximum primary energy is 10$^{21.5}$ eV in both cases. For proton primaries (left) two peaks are apparent in the electron neutrino fluxes: the peak at lower energies corresponds to electron neutrinos produced in the decay of neutrons arising from
 photopion interactions, while the one at higher energies correspond to electron neutrinos coming from the decay of muons produced by pions.  Since electron neutrinos produced by neutron decay  carry $\sim 4 \times 10^{-4}$ of the neutron energy, the maximum neutrino  energy from neutron decay in $\sim 10^{18}$ eV, while the lowest energy is $\sim 2 \times 10^{16}$ eV,  corresponding to  primaries at the proton photopion production threshold, $E_{\gamma \pi} \simeq 6 \times 10^{19}$ eV. Charge pion decays produce three neutrinos (2 muon neutrinos and 1 electron neutrino) each carrying  $\sim$ 5\%  of the nucleon energy. Therefore, the energy of neutrinos from charged pion decay ranges from  $\sim 2 \times 10^{20}$ eV  down to  $\sim 3 \times 10^{18}$ eV. The average number of photopion interactions that a 10$^{21.5}$ eV proton has as it traverses 300 Mpc is $\sim 18$.

The right panel in Fig. \ref{dndloge} shows the fluxes of electron and muon neutrinos arising from the propagation of iron nuclei with energy 10$^{21.5}$ eV over a distance of 300 Mpc. Three curves are displayed: (1)  the flux of  electron neutrinos arising from the decay of neutrons from the photodisintegration of nuclei (dotted line); (2)  the flux of electron neutrinos from photopion interactions of secondary nucleons (solid line), which shows the double two peaks, one from the decay of neutrons and the other from the decay of muons;  (3)  the flux of muon neutrinos
 from photopion interactions (dashed line). For iron at 10$^{21.5}$ eV, the secondary nucleons emitted by photodisintegration have on average $\sim 6 \times 10^{19}$ eV,  which is  just around the threshold for photopion production. Thus, secondary nucleons from iron at 10$^{21.5}$ eV have only about one photopion interaction on average, and a narrow peak of electron and muon neutrinos from the decay of the pions should be around $3 \times 10^{18}$ eV.  Similarly, the spectrum of electron neutrinos from neutron decay should peak 
 $\sim 2 \times 10^{16}$ eV). These estimates are in good agreement with the results from our  simulations. Note that the spectrum of neutrinos is less broad in the case of iron primaries compared to protons, as most of the neutrinos arise from the first interaction of secondary nucleons  which have a  small dispersion in the energy. The amplitude of the iron fluxes may seem high compared to the proton case, but the small number of photopion interactions in the case of iron is compensated by  56 nucleons that can interact once.

\section{Neutrinos from a cosmological distribution of sources}

The neutrino spectrum produced by protons or iron from a single source  is fully developed if the source distance to the observer is greater than about 300 Mpc. We calculate the neutrino flux for a distribution of sources by integrating the single source results over redshift with the appropriate redshift  dependence and with a homogeneous source distribution as in \cite{ESS01}. We further assume that sources have the same primary injection spectrum.

The local neutrino flux $ {\cal F}_i (E_{\nu_i})$ of neutrino flavor $\nu_i$ generated by the propagation of cosmic rays over cosmological distances can be written as an integral 
over redshift and primary energy at the source $E_p^s$.
\begin{equation}
 {\cal F}_i (E_{\nu_i}) =  \int\!\!
 \int \frac{Y(E_p^s,E^s_{\nu_i},z)}{{4 \pi} E_\nu^s} \frac{dN^s_p} {dE^s_{p}}
 \frac{dV_c}{dz} \frac{n_c(z)}{4\pi d^2_L} dz dE_p  \ ;
\label{neuflux}
\end{equation}
where ${dN^s_p} /{dE^s_{p}}$ is the number of primaries per unit
 time and energy produced by the one source, $n_c(z)$ is the number of
 sources per comoving volume, $V_c$ is the comoving volume, $d_L$ is
 the luminosity distance and $Y(E^s_p,E^s_{\nu_i},z)$=$E^s_{\nu_i}
({dN_{\nu_i}}/{dN_p dE^s_{\nu_i}})$ is the neutrino yield. The
 dependence of the neutrino yield on redshift arises from the presence
 of a higher cosmic background temperature at higher redshift.

Eq. \ref{neuflux} can be simplified using:
\begin{equation}
 \frac{dV_c}{dz} \frac{1}{4\pi d^2_L}=\eta(z) \frac{c}{1+z} \ ;
\label{comeq}
\end{equation}
where $\eta(z)$ is  given by:
\begin{eqnarray}
\eta (z) &=& \frac{dt}{dz} = {{1} \over {H_0 (1 + z)}}
\big[\Omega_M ( 1 + z)^3 + \Omega_\Lambda
\nonumber\\
& & + (1-\Omega_M-\Omega_\Lambda) (1+z)^2\big]^{-1/2} \ .
\label{eta-factor}
\end{eqnarray}
 We use cosmological parameters from  \cite{wmap}. 

The density of sources per comoving volume can be expressed as:
\begin{equation}
n_c(z)=n_0~{\cal H}(z) ,
\end{equation}
where $n_0$ is the density of sources at $z$=0 and ${\cal H}(z)$ express a 
 evolution of sources with redshift.

 To express the integrand of Eq.\ref{neuflux} in terms of the ``observed''
 neutrino energy, we  redshift the energy  ($(1+z) E^s_{\nu_i}$=$E_{\nu_i}$) and the bandwitdh ($(1+z) dE^s_{\nu_i}$=$dE_{\nu_i}$).  In addition,  $E_\nu$ and $E_p^s$ scale with $(1+z)$ maintaining the
same invariant reaction energies in the presence of a higher cosmic
background temperature.

Finally, the neutrino yield, $Y$, is evaluated using the Monte Carlo result for a 300 Mpc source and the scaling relation: 
\begin{equation}
Y(E_p^s, E_\nu,z)  = Y( (1+z) E_p^s, (1+z)^2 E_\nu,z=0).
\end{equation}
As a result, Eq. \ref{neuflux} can be expressed as:
\begin{eqnarray}
 {\cal F}_i (E_{\nu_i})  &=& \frac{c}{4 \pi E_{\nu_i}} \int\!\!
 \int Y( (1+z) E_p^s,(1+z)^2 E_{\nu_i},0) 
 \nonumber\\
  & &  {\cal H}(z) \eta(z) {\cal L}_0(E_p^s) 
\frac{dE^s_p}{E^s_p}  dz  \ .
\label{flux2}
\end{eqnarray}
where ${\cal L}_0$ is the number of primaries injected per unit of volume, time, and $dlogE^s_p$  at $z$=0, i.e., 
\begin{equation}
{\cal L}_0(E_p^s) = n_0 E^s_p \frac{dN^s_p} {dE^s_{p}} \ .
\end{equation}
${\cal L}_0$ is related to the injection power required to maintain 
 the present cosmic ray density per unit volume and time,
\begin{equation}
 P_0 = n_0 \int_{E_{\rm min}}^{E_{\rm max}} E^s_p \frac{dN_p}{dE_p^s} dE_p^s  \ .
\label{lumino}
\end{equation}

We assume a primary injection spectrum given by
\begin{equation} 
 \frac{dN_p}{dE_p^s} = \left\{
\begin{array}{ll}
\beta~E^{-\alpha} &, E < E_{\rm max}  \\
  0  &, E > E_{\rm max} \\
\end{array}\right.
\label{inj-spectrum}
\end{equation} 
where $\beta$ is a constant and we fixed the spectral index $\alpha$=2 and the maximum energy varied up to $E_{\rm max} = Z \  4 \times 10^{20}$ eV, i.e.,  $10^{22}$ eV for iron nuclei.

For the cosmological evolution of cosmic ray sources, ${\cal H}(z)$,  we use the parametrization
of \cite{Waxman95}:
\begin{equation}
  {\cal H}(z) =  \left\{
\begin{array}{lcl}
 (1 + z )^n  &  {\rm :}   &  z <  1.9\\
 (1 + 1.9)^n &  {\rm :}   & 1.9 < z < 2.7\\
 (1 + 1.9)^n e^{\frac{(2.7-z)}{2.7}}& {\rm :} &
  z >  2.7 \ .
\end{array}\right. 
\label{evo} 
\end {equation}
First we considered a mild redshift evolution with $n=3$ in Eq.~\ref{evo}.  For this case, the  $\eta(z){\cal H}(z)$ term in Eq.~\ref{flux2} evolves approximately as $(1+z)^{0.5}$ for
$z<$1.9 ,  as $(1+z)^{-5/2}$ between $z$=1.9 and 2.7, and exponential decreases for 
$z \geq 2.7$.  We later consider a model with stronger evolution given by 
$n=4$ up to $z=1.9$, followed by a constant $ {\cal H}(z)$ for $z \geq 1.9$.

As discussed in \cite{ESS01}, the redshift scaling of the neutrino yield is not exact. The photopion production of neutrinos is slightly overestimated for redshifts $\lesssim 0.08$ compared to Fig.~\ref{dndloge}, but the total contribution from low redshifts is relatively small. At high redshifts, neutron decay becomes subdominant to photopion production, thus, we expect a modest overestimate of the $\bar{\nu}_e$ flux around $10^{16}$~eV. In addition, the sum of $\bar{\nu}_e$ and $\nu_e$
 fluxes at high energies remains unchanged, but the flavor distribution may be altered.
The most uncertain effect in the case of iron primaries is the evolution of the IRB. The IRB may have a more complex redshift evolution due to the density and/or luminosity evolution of infra-red sources such as galaxies and active galactic nuclei (AGN). The IRB evolution influences the neutrino yield through its effect in the photodisintegration.  However, for the energies we consider, the disintegration is efficient even if the background is lowered by an order of magnitude.

Another necessary ingredient to determine the neutrino flux is the normalization of the injected cosmic ray flux or the injection rate $P_0$. A number of alternative estimates can be used \cite{ESS01} which depend on assumptions about the unknown energy range  where a transition from galactic to extra-galactic cosmic rays occurs,  how the propagation effects on the injection spectrum, and the limiting energies of extragalactic accelerators. We chose to  calculate the product $\beta n_0$ in Eq. \ref{lumino} integrating between 10$^{19}$ eV and 10$^{21}$ eV with $\alpha$=2, and assuming the value of $P_0  = 4.5 \times {\rm 10}^{44}$ erg/Mpc$^3$/yr  \cite{Waxman95}. (Note that $P_0$ in \cite{Waxman95} was calculated for a evolution of sources given by $(1+z)^3$.)
 Once we set $\beta n_0$, different primary spectra can be assumed without changing the normalization at low energies. 
 
\begin{figure}
\epsfig{file=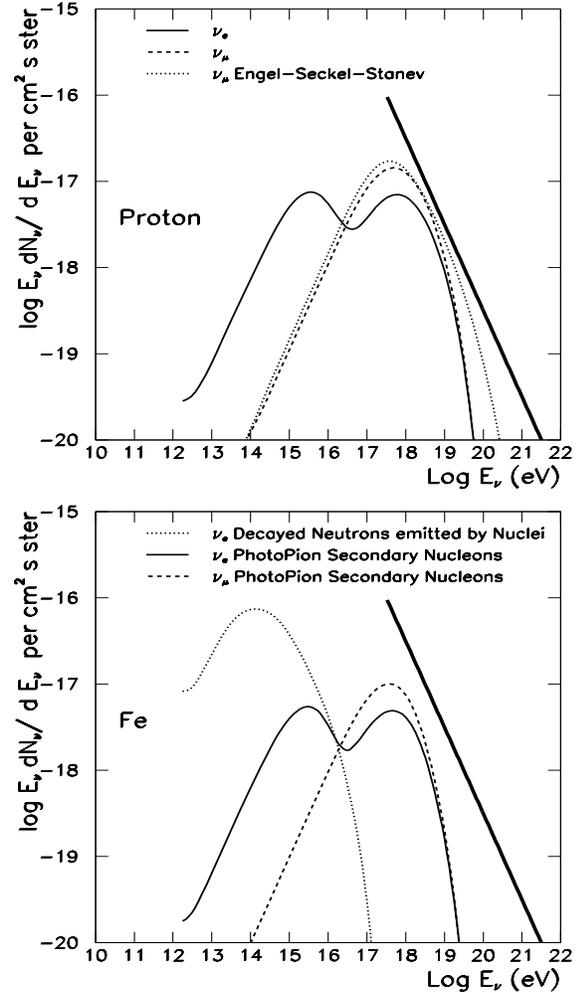,width=1.0\linewidth ,height=1.7\linewidth}
\caption{Electron and muon neutrino fluxes obtained with our nominal choice of astrophysical and cosmological parameters. Left (right) graph shows the results for proton (iron) primaries with a maximum energy at acceleration of $4~Z\times10^{20}$ eV. The  labels are the same as Fig. \ref{dndloge}. We also show the proton flux obtained by \cite{ESS01} by a dotted line in the proton graph. We show the WB limit \cite{WB1,WB2} by a thick solid line in both plots.}
\label{nuflux}
\end{figure}

Figure~\ref{nuflux} shows electron and muon neutrino fluxes obtained with our nominal choice of astrophysical and cosmological parameters ($n=3$, $\alpha=2$), and an integration to redshift of $z_{\rm max} = 8$. Integrating to infinity increases the neutrino fluxes only by about 5\%. Left (right) graph shows the results for proton (iron) primaries with a maximum energy at acceleration of $4~Z\times10^{20}$ eV. The same neutrino components are displayed as in Fig. \ref{dndloge}.

For comparison, in the proton panel of  Fig.~\ref{nuflux},  we show the neutrino flux from proton propagation as obtained by  \cite{ESS01}. The agreement between  \cite{ESS01} and our results is quite good. The small difference in the peak of $\nu_\mu$ is likely due to multi-pion production that we did not consider in this calculation.  At higher energies ($\sim 10^{19}$ eV) our flux is lower because our maximum energy at acceleration is lower.

	We also show in Fig. \ref{nuflux}  the maximum neutrino flux produced by cosmic ray sources derived by Waxman \& Bahcall~\cite{WB1,WB2} (WB) for the same assumptions including source evolution model,  spectra and $P_0$. The WB flux assumes a maximally thin source with energy input into protons equal to neutrinos (for a discussion of these assumptions and alternative models, see e.g., \cite{kalashev02}).  The comparison shows that  around $10^{18} - 10^{19}$ eV, neutrinos produced by propagation can become comparable to the maximum production at the source. At lower energies, propagation is not as effective in generating neutrinos as sources may be since photopion production stops.

The left and right panels of Fig. \ref{nuflux} show that neutrino fluxes from proton and iron propagation are quite similar, even if the detailed neutrino generation processes differ. An important new feature of the iron generated neutrino flux is the  peak at low energies ($\sim 10^{14}$ eV)  which corresponds to the decay of the neutrons emitted by the disintegrating nuclei.   
The position of the peak can be understood in terms of the threshold energy for photodisintegration of iron, $E_{pd}$ which is $ \sim 10^{20}$ eV at $z$=0  and decreases with redshift.  Nuclei are completely photodisintegrated at energies larger than $ 10^{20.5}$ eV $/(1+z)$.
Since the primary spectrum is steep, most neutrinos are produced  by primaries at the threshold for photodisintegration.  Iron nuclei at the threshold of photodisintegration produce neutrinos with an energy $4~10^{-4} E_{pd}/A \simeq  7 \times 10^{14}$ eV at $z$=0.  For larger redshifts this energy is lower because the threshold of photodisintegration is reduced but also because the neutrino energy we observe is redshifted from the neutrino energy at production. At $z$=3 , the neutrino energy at photodisintegration threshold  is   $\sim 4 \times10^{13}$ eV. Contributions from larger redshifts are suppressed by the exponential decrease of sources. Therefore, it  is clear why the neutrino flux peaks at an energy of $10^{14}$ eV.

The neutrino fluxes from iron and proton primaries at higher energies have the same basic origin: photopion interactions. The threshold for photopion production at  $z$=0 is $E_{\gamma \pi} \simeq 6 \times 10^{19}$ eV for protons and $\sim 3 \times 10^{21}$  eV for
iron primaries. This threshold energy decreases as $z$ increases due to the increasing cosmic background temperatures. As the threshold decreases, the more abundant lower energy primaries can contribute.

Below $10^{16}$ eV the neutrino flux is dominated by electron neutrinos. In the iron case,
there are two sources for these neutrinos: neutrinos from the decay of neutrons emitted by the nuclei and neutrinos from secondaries emitted in photopion interactions.
For $z=0$, the energy of neutrons that can produce these neutrinos by decay should be greater than  $\sim 56 \times 10^{16}~{\rm eV}/4\times10^{-4} \sim 1.4 \times 10^{21}$  eV.  At larger redshifts, the neutrino energy redshifts and a larger primary neutron energy is required to produce 10$^{16}$ eV neutrinos.  At $z=0$, iron nuclei need to have  energies larger than 10$^{21}$ eV to allow secondary nuclei to produce neutrinos via photopion interactions. For larger redshifts, this threshold decreases because the larger background photon temperature, allowing contributions from smaller primary energies. At redshifts larger than $z=3$ the flux is suppressed by the exponential decrease in the source distribution.

In the proton case, all neutrinos originate in photopion interactions, either through the decay of neutrons or pions. The contribution at $E_{\nu} \simeq 10^{16}$ eV is maximized by pion decay at $z$=2 and $E_p \sim 10^{20}$ eV. This is similar  to the photopion generation in the iron case, but  with a shift in the primary energies that contribute to the flux to lower energies. The similarity between the iron and proton fluxes can be inferred from Fig. \ref{numneu}: the higher flux at proton threshold is compensated by the larger number of neutrinos produced at the iron threshold.

Neutrinos with energies larger than $10^{16}$ eV must be produced by primaries with increasingly higher energies.  For example, neutrinos with an energy of $10^{19}$ eV  are produced by primary protons (iron) with energy $\gtrsim 10^{20.5}$ eV  ($\gtrsim10^{21.5}$ eV) at $z$=0 (note that, in the proton case, the primary energy is close to the maximum energy at acceleration). At higher redshifts the contribution to the neutrino flux at $10^{19}$ eV  is reduced by low number of primaries combined with the redshift of the neutrino energy ($(1+z)^{-1}$) and the contribution of $\eta(z){\cal H}(z)$$\sim (1+z)^{0.5}$. As a consequence, the high  energy part of the neutrino spectra is dominated by nearby sources.

\begin{figure}
\epsfig{file=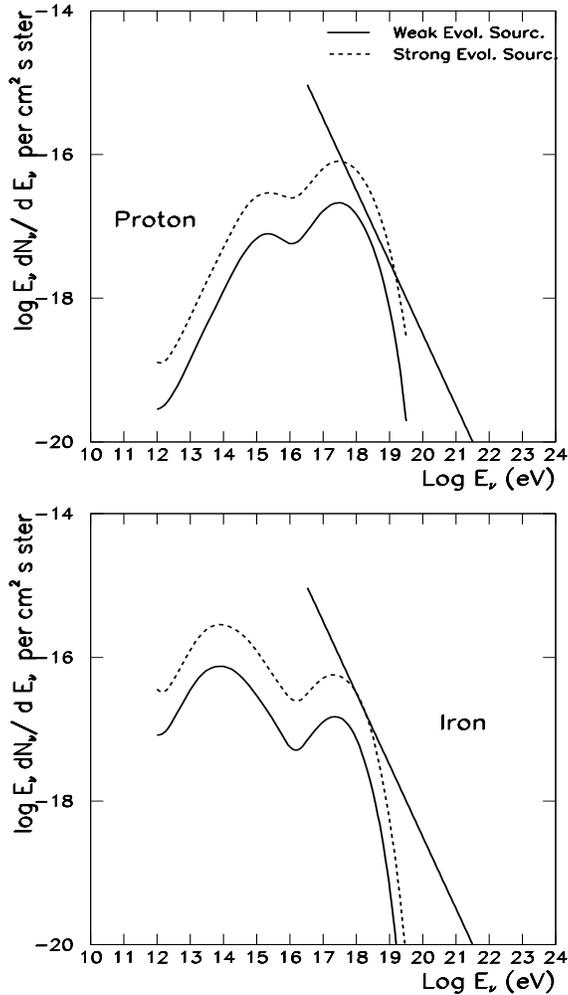,width=1.0\linewidth ,height=1.7\linewidth}
\caption{Dependence of neutrino fluxes on the cosmological evolution of  sources. The left (right) graph correspond to the iron (proton) sources. Continuous (dashed) line correspond to $(1+z)^3$ ($(1+z)^4)$) evolution up to $z$=1.9, constant between 1.9 and 2.7, and
 beyond 2.7 exponentially suppressed (constant).}
\label{nuflux_strong}
\end{figure}

The proton and iron generated neutrino fluxes in Fig. ~\ref{nuflux} are strongly dependent on the choice of parameters for the ultra-high energy cosmic ray sources, such as  the redshift evolution and source spectrum (slope, amplitude, and maximum energy). The cosmological evolution of the source luminosity  could be  stronger than the $n=3$ choice is Eq.\ref{evo}, such as estimates for the evolution of star formation ~\cite{Madau} and  gamma-ray bursts ~\cite{GRB_evo,Pugliese}.  In Fig.~\ref{nuflux_strong}, we show the neutrino flux with the same parameters  as Fig.~\ref{nuflux}, but a stronger cosmological evolution: with $n=4$ in Eq.~\ref{evo} up to $z=1.9$ and constant for $z \geq 1.9$.  The stronger cosmological evolution increases the neutrino flux by  a factor 
of $\sim$ 3 and generates a small shift of the maximum flux  to lower energy. Again left (right) graphs correspond to the  proton (iron) case. 

\begin{figure}
\epsfig{file=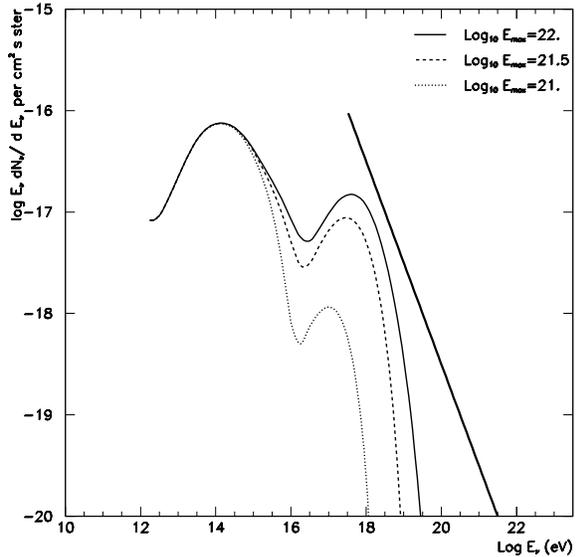,width=1.0\linewidth ,height=1.0\linewidth}
\caption{Dependence of neutrino fluxes from  iron primaries on $E_{\rm max}$. $E_{\rm max} =10^{21}$eV (dotted), $10^{21.5}$eV (dashed),  and $10^{22}$ eV (solid) with different power injection rate  keeping same cosmic ray flux at $10^{19}$ eV.}
\label{nuflux_cutoff}
\end{figure}

The neutrino flux produced by the propagation of heavy nuclei is extremely sensitive to the
 maximum cosmic-ray energy at the source, $E_{\rm max}$.  Current experiments indicate that observed cosmic rays can reach energies significantly higher than $10^{20}$ eV.  In Fig. \ref{nuflux_cutoff},  we consider  $E_{\rm max} =10^{21}$eV, $10^{21.5}$eV,  and $10^{22}$ eV, and a different power injection rate is used for each maximum energy to maintain the normalization of the injected spectra at $10^{19}$ eV. The resulting fluxes in Fig. \ref{nuflux_cutoff} show the significance of having $E_{\rm max} \gtrsim 10^{21.5}$ eV: the flux is an order of magnitude lower if $E_{\rm max} = 10^{21}$ eV and a factor of 3 larger if $E_{\rm max}=10^{22}$ eV.

 Iron primaries are more easily accelerated than protons because of their reduced rigidities. However, during acceleration the interaction of iron primaries with the infrared background near the source may limit the maximum energy more strictly than the confinement requirements usually applied to limit proton acceleration. The lifetime of an iron nucleus at this energies is about $\tau = 2\times 10^{14}$ s due to photodisintegration in the CMB and IRB.

\section{Conclusions}

We have calculated the neutrino fluxes associated with the propagation of cosmic rays from extragalactic sources of proton and iron primaries. The propagation of iron produces a new flux of neutrinos from neutron decay at energies $\sim 10^{14}$ eV. At higher energies the neutrino flux from proton and iron propagation have similar features and have comparable amplitudes if the maximum energy at acceleration is scaled by rigidity, i.e.,  $E_{\rm max} (Z) \propto Z$.  The peak flux at high energies is reached at a lower energy in the case of iron when compared with proton propagation. The resulting neutrino flux is most sensitive to the choice of maximum energy at acceleration followed by the cosmological evolution of sources. 

In conclusion, if  cosmic rays at the highest energies are produced by acceleration mechanisms
 in astrophysical sources distributed homogeneously throughout the universe, and the maximum energy at acceleration is $\gtrsim Z \times 10^{20.5}$, significant neutrino fluxes will be produced regardless of the
 primary composition. For reasonable choices of source parameters, the rate of neutrino events in future high energy neutrino detectors, such as IceCube, Auger, EUSO, ANITA, and SALSA, from iron primaries is lower by a factor of a few when compared with the proton generated flux. However, the sensitivity of the resulting flux to cosmic ray source parameters, shows that rates can be lower by an order of magnitude.

\section*{Acknowledgements}
This work was supported in part by the KICP under NSF PHY-0114422 ,
 by  the NSF through grant AST-0071235, and the DOE grant DE-FG0291-ER40606 at the University of Chicago. A.A.W. acknowledges the support of PPARC, UK and the hospitality of the KICP in Chicago. 
 
 After we presented this work in \cite{cris04,leeds04}, we learned of a similar calculation \cite{hooper04}. We find some similarities and disagreements between the two calculations. Their  iron flux is smaller than ours by an order of magnitude. The choice of an exponential cutoff versus a sharp cutoff may explain some of the differences, but the exact source of the full disparity is not clear. Only a small fraction of the iron generated flux is displayed in \cite{hooper04} and the detailed description of their calculation procedure is lacking. In particular, we expect the peak flux for iron to move to lower energies and to be a bit narrower when compared with that generated from  protons as we see in Fig. \ref{dndloge}.

\end{document}